\documentclass[prl,letterpaper,aps,floatfix,twocolumn]{revtex4-1}
\usepackage{graphicx}
\usepackage{amsmath}
\usepackage{subfigure}
\newcommand{\beq}{\begin{equation}}
\newcommand{\eeq}{\end{equation}}
\newcommand{\bk}{{{\bf{k}}}}

\newcommand{\ba}{{\bf{a}}}

\newcommand{\bb}{{\bf{b}}}

\newcommand{\beqa}{\begin{eqnarray}}
\newcommand{\eeqa}{\end{eqnarray}}

\newcommand{\bGamma}{{\boldsymbol \Gamma}}

\begin{document}
\title{Mirror Anomaly in Dirac Semimetals}
\author{A.A. Burkov}
\affiliation{Department of Physics and Astronomy, University of Waterloo, Waterloo, Ontario 
N2L 3G1, Canada} 
\date{\today}
\begin{abstract}
We demonstrate that, apart from the chiral anomaly, Dirac semimetals possess another quantum anomaly, 
which we call the mirror anomaly, and which manifests in a singular response of the Dirac semimetal to an applied magnetic field. 
Namely, the anomalous Hall conductivity exhibits step-function singularities when the field is rotated. We show that this phenomenon 
is closely analogous to the parity anomaly of $(2+1)$-dimensional Dirac fermions, but with mirror symmetry, which we demonstrate emerges near any Dirac point at a time reversal invariant momentum, replacing the parity symmetry.  
\end{abstract}
\maketitle
Response of topologically nontrivial states of matter, both insulating and metallic, to external fields, may often 
be understood in terms of quantum anomalies~\cite{Haldane88,Volovik03,Volovik07,Furusaki13,Werner13,Burkov_ARCMP}, which is a concept that originated in particle physics~\cite{Adler69,Jackiw69}, but has now found its way into condensed matter. 
Anomaly in the particle physics context refers to violation of a ``classical" symmetry, i.e. symmetry of the Lagrangian,
once the second quantization is performed. 
For example, perhaps the most well known of quantum anomalies, the chiral anomaly, arises due to violation of the chiral symmetry 
of a massless Dirac Lagrangian. 

It is well established~\cite{Spivak12,Burkov_lmr_prl,Parameswaran14,Burkov_lmr_prb,Burkov_AHE,Ong_anomaly,Li_anomaly}
 that the chiral anomaly leads to observable manifestations in condensed matter realizations 
of massless chiral particles, i.e. in Dirac and Weyl semimetals~\cite{Wan11,Ran11,Burkov11-1,Burkov11-2,Xu11,Kane12,Fang12,Fang13,Chen14,Neupane14,HasanTaAs,DingTaAs2,DingTaAs,Lu15,Weyl_RMP,Hasan_ARCMP,Felser_ARCMP}.
In particular, both the anomalous Hall effect in magnetic Weyl semimetals and the negative longitudinal magnetoresistance, or the chiral magnetic effect~\cite{Kharzeev08}, in both Weyl and Dirac semimetals may be understood as being a consequence of the chiral anomaly. 

In this paper we demonstrate that Dirac semimetals also possess another kind of anomaly, which is distinct from the chiral anomaly
and is instead closely related to the parity anomaly of $(2+1)$-dimensional relativistic fermions. 
Parity anomaly in the relativistic context refers to violation of the parity (and time reversal) symmetry of a massless 
$(2+1)$-dimensional Dirac Lagrangian when second quantization in the presence of electromagnetic fields is performed: a topological 
Chern-Simons term, violating parity and time reversal, is generated in the action when the fermions are integrated out~\cite{Semenoff84}. 
In condensed matter realizations of two dimensional (2D) Dirac fermions the parity anomaly has a somewhat different, but closely related, meaning, referring to a singular step function dependence of the anomalous Hall response of a 
massive (i.e. gapped) 2D Dirac fermion on the gap magnitude and sign~\cite{Haldane88}.  
Here we show that something very similar happens in three dimensional (3D) Dirac semimetals, but with mirror symmetry, which emerges near any Dirac point at a time reversal invariant momentum (TRIM), replacing parity. 
We thus call the corresponding anomaly the mirror anomaly. 
We demonstrate that it manifests in a singular response of the Dirac semimetal to an applied magnetic field and should be readily observable experimentally. 

We will restrict ourselves to a particular kind of Dirac semimetal, with a single (or several symmetry-related) 
Dirac point at a TRIM (type-II Dirac semimetal~\cite{Kane12,Weyl_RMP}). 
Analogous effects should, however, also exist in the second type of Dirac semimetal, with two Dirac points on a rotation axis 
(type-I Dirac semimetal)~\cite{Fang12,Fang13}.
Experimental realizations of type-II Dirac semimetals include TlBi(S$_{1-x}$ Se$_x$)$_2$~\cite{Ando11},
(Bi$_{1-x}$In$_x$)$_2$Se$_3$~\cite{Brahlek12}, and ZrTe$_5$~\cite{Li_anomaly,Chen15}.

A minimal model of a Dirac band-touching point at TRIM in 3D involves four degrees of freedom 
per unit cell: two orbital and two spin.
We introduce two sets of Pauli matrices $\tau^i$ and $\sigma^i$, which will represent operators acting on the orbital and 
spin degrees of freedom correspondingly. 
In the presence of inversion symmetry, the two orbital states may always be chosen to be related to each other 
by the parity operator $P$. 
We thus take the orbital states to be the eigenstates of $\tau^z$, in which case the parity operator $P = \tau^x$. 

The most general time reversal and parity invariant momentum space Hamiltonian, describing the above system, may be 
written as~\cite{Fu-Kane}
\beq
\label{eq:1}
H(\bk) = d_0(\bk)  + \sum_{a = 1}^5 d_a(\bk) \Gamma^a, 
\eeq
where $\Gamma^a$ are the five matrices, realizing the Clifford algebra $\{\Gamma^a, \Gamma^b \} = 2 \delta^{ab}$, even under the product of parity and time reversal 
$P T$. 
Explicitly, the five $\Gamma$-matrices are given by
\beq
\label{eq:2}
\Gamma^1 = \tau^x,\,\, \Gamma^2 = \tau^y,\,\, \Gamma^3 = \tau^z \sigma^x,\,\, \Gamma^4 = \tau^z \sigma^y,\,\,
\Gamma^5 = \tau^z \sigma^z, 
\eeq
where $\Gamma^1$ is even under both parity and time reversal, while $\Gamma^{2-5}$ are odd under both separately, 
but even under their product. 

Suppose Dirac band touching points are realized at crystal symmetry related set of TRIM. 
For simplicity, let us take the TRIM to be at the $\Gamma$-point in the first Brillouin zone (BZ), in
which case this is a single point (such a Dirac node may only arise through fine tuning and can not be 
crystal symmetry protected~\cite{Kane12}). Generalization to the case of multiple symmetry related TRIM is straightforward 
and the results do not change in a qualitative way. 
Taylor expansion of Eq.~\eqref{eq:1} near the $\Gamma$-point will have the following general form, in which 
we use relativistic Dirac matrix notation
\beq
\label{eq:3}
H(\bk) \approx (\alpha + \beta \gamma^0) \bk^2 + v_F \gamma^0 (\gamma^1 k_x + \gamma^2 k_y + \gamma^3 k_z),
\eeq
where $\alpha$, $\beta$ and $v_F$ are expansion coefficients and we have assumed cubic symmetry for notational simplicity 
(this does not affect any of the arguments below).
We will ise $\hbar = 1$ units throughout, except in the final results. 
Since $\Gamma^1$ is the only parity and time reversal invariant $\Gamma$-matrix, we have $\gamma^0 = \Gamma^1 = \tau^x$. 
The other three matrix coefficients in Eq.~\eqref{eq:3}, $\gamma^0 \gamma^i$, $i = 1,2,3$, may in general be given by 
any three independent linear combinations of the $\Gamma^{2-5}$ matrices. 

Let us view Eq.~\eqref{eq:1} as a lattice Fourier transform of a tight-binding Hamiltonian. 
Then it is clear that the physical origin of the term, proportional to $\Gamma^2$, is spin-independent hopping between the two orbital states, labelled by the eigenvalues of $\tau^z$, and located in different (e.g. neighboring) unit cells, 
while the physical origin of the terms, proportional to $\Gamma^{3,4,5}$ are spin-dependent hopping terms, which arise due to the spin-orbit interactions. 
It follows that for any physical realization of a Dirac semimetal, the $\Gamma^2 = \tau^y$ matrix will always be 
present in the Taylor expansion of the Hamiltonian Eq.~\eqref{eq:3} at linear order. 
Indeed, restricting the spin-independent hopping to nearest neighbor sites for simplicity, the coefficient $d_2(\bk)$ of the 
matrix $\Gamma^2$ in Eq.~\eqref{eq:1} has the following general form
\beq
\label{eq:4}
d_2(\bk)  = - t_1\sin(\bk \cdot \ba_1) - t_2 \sin(\bk \cdot \ba_2) - t_3 \sin(\bk \cdot \ba_3), 
\eeq
where $\ba_{1,2,3}$ are the primitive translation vectors of the Bravais lattice and $t_{1,2,3}$ are the 
hopping amplitudes, corresponding to the directions $\ba_{1,2,3}$. 
Expanding to linear order near an arbitrary TRIM $\bGamma$, we then obtain
\beq
\label{eq:5}
d_2(\bGamma + \delta \bk) \approx - \delta \bk \cdot \sum_{i = 1}^3 t_i \ba_i \cos(\bGamma \cdot \ba_i).
\eeq  

TRIM may generally be written as half a reciprocal lattice vector $\bGamma = (m_1 \bb_1 + m_2 \bb_2 + m_3 \bb_3)/2$,
where $\bb_i$ are primitive translation vectors of the reciprocal lattice and $m_i$ are integers. 
Then $\cos(\bGamma \cdot \ba_i) = \cos(\pi m_i) = (-1)^{m_i}$. 
It then follows from the linear independence of the primitive translation vectors $\ba_i$ that the linear term in the Taylor expansion of $d_2(\bk)$ near any TRIM $\bGamma$ is always nonvanishing. 

To understand the consequences of this, let us consider the chirality operator $\gamma^5 = i \gamma^0 \gamma^1 \gamma^2 \gamma^3$. It is even under time reversal but odd under parity and thus must have the following general form
\beq
\label{eq:8}
\gamma^5 = \alpha \tau^z + \beta_i \tau^y \sigma^i, 
\eeq
where $i = x,y,z$. 
By rotating spin quantization axes, we may always bring Eq.~\eqref{eq:8} to the following form
\beq
\label{eq:9}
\gamma^5 = \alpha \tau^z + \beta \tau^y \sigma^z. 
\eeq
The property that the $\Gamma^2 = \tau^y$ matrix is always present at linear order in the Taylor expansion of the Hamiltonian 
implies that the coefficient $\beta$ in Eq.~\eqref{eq:9} is always nonzero, i.e. $\gamma^5$ always involves one of the spin components. 
Indeed, $\gamma^5$ can only be spin-independent if, up to all possible permutations of $x,y$ and $z$, 
$\gamma^1 \propto \sigma^x$, $\gamma^2 \propto \sigma^y$, $\gamma^3 \propto \sigma^z$. 
This, however, is impossible, as demonstrated above, since at least one of the $\gamma^i$ must involve a spin-independent 
contribution. 
Eq.~\eqref{eq:9} implies that only one of the spin components commutes with $\gamma^5$, while the other two do not. 
This has significant consequences for the Zeeman response of the Dirac semimetal to an applied magnetic field, as will be demonstrated below. 

To proceed, let us consider a specific Dirac Hamiltonian, satisfying the properties, described above. 
Let us assume that we have a single Dirac point at $\bk = 0$ and the Taylor expansion of the functions $d_a(\bk)$ 
near this point has the following form
\beqa
\label{eq:10}
&&d_1(\bk) \approx \frac{\Delta \bk^2}{2}, \,\, d_2(\bk) \approx v_F k_z, \,\, d_3(\bk) \approx v_F k_y, \nonumber \\
&&d_4(\bk) \approx - v_F k_x, \,\, d_5(\bk) \approx \frac{\lambda}{2} (k_z^2 - k_x^2) k_y,
\eeqa
which arises, e.g. in the Fu-Kane-Mele diamond lattice model~\cite{Fu-Kane}. 
We have assumed the Fermi velocities to be the same in all directions for simplicity and have also taken $d_0(\bk)=0$, as it will not affect any of the arguments or the final results in a significant way. 
This corresponds to the following representation of the Dirac gamma-matrices:
\beq
\label{eq:11}
\gamma^0 = \tau^x, \,\, \gamma^1 = i \tau^y \sigma^y, \,\, \gamma^2 = - i \tau^y \sigma^x, \,\, \gamma^3 = i \tau^z, \,\, 
\gamma^5 = \tau^y \sigma^z. 
\eeq 
We now note the following property of the Dirac Hamiltonian, which will play a crucial role in what follows. 
Namely, the linearized Dirac Hamiltonian 
\beq
\label{eq:12}
H({\bk}) = v_F (- \tau^z \sigma^y k_x + \tau^z \sigma^x k_y + \tau^y k_z), 
\eeq
commutes with the operator $M = i \sigma^x$ in the $k_x = 0$ plane
\beq
\label{eq:13}
[H(k_x = 0), M] = 0. 
\eeq
To understand the physical meaning of the operator $M$ we recall the connection between the gamma-matrices and generators 
of rotations. 
In particular, the generator of rotations about the $x$-axis is given by 
\beq
\label{eq:14}
\sigma^{32} = \frac{i}{2} [\gamma^3, \gamma^2] = \tau^x \sigma^x, 
\eeq
which implies $M$ is the operator of reflection in the yz ($k_x = 0$) plane
\beq
\label{eq:15}
M = P R^x_{\pi} = i \gamma^0 \sigma^{32} = i \sigma^x. 
\eeq
Thus we come to the conclusion that the linearized Dirac Hamiltonian Eq.~\eqref{eq:12} possesses mirror symmetry 
in the $yz$-plane (same is true of the $xz$-plane as well of course). 
The cubic term $d_5(\bk) \tau^z \sigma^z$ violates this mirror symmetry and thus the symmetry is only an approximate 
low-energy symmetry (it may be an exact crystalline symmetry as well, but in general is not). 
We will see shortly that this emergent mirror symmetry of the Dirac point leads to important observable consequences for the magnetic response of Dirac semimetals. 

Let us now assume that an external magnetic field is applied to the Dirac semimetal, which may be rotated in any direction. 
For concreteness, let us assume that the field is rotated in the $xz$-plane, and consider the anomalous Hall conductivity 
in the $xy$-plane, $\sigma_{xy}$, as a function of the angle $\theta$ of the field with respect to the $z$-axis. 
By anomalous Hall conductivity we mean here the part of the total Hall conductivity, which arises from the Zeeman splitting effect of the magnetic field, as opposed to the orbital effect (the Lorentz force). The two may be separated experimentally in the standard way by subtracting off the high-field linear part of the Hall resistivity. 

In order to understand what happens as the field is rotated, we first note that while $\sigma^z$ commutes with the chirality operator 
$\gamma^5 = \tau^y \sigma^z$, $\sigma^{x,y}$ do not. This is a general property of Dirac semimetals, as explained above. 
As a consequence of this, the applied magnetic field will have a very different effect on the spectrum, depending on its direction: 
while the field, directed along the $z$-axis, will split the Dirac node into a pair of Weyl nodes (as it conserves the chiral charge), the field along the $x$ or $y$-direction will have a more complex effect. 

\begin{figure}[t]
\includegraphics[width=8cm]{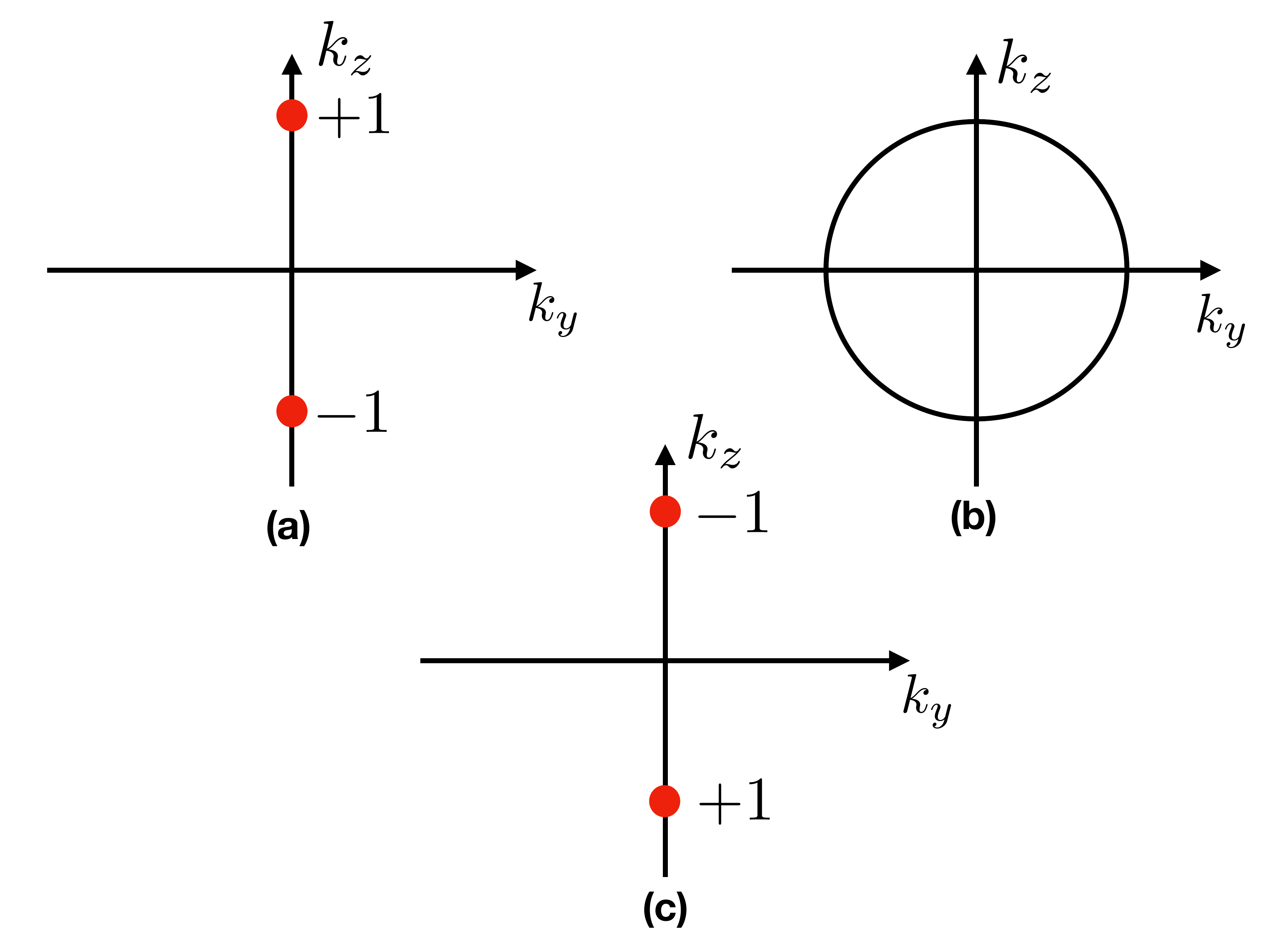}
\caption{(Color online) Evolution of the Weyl nodes in the presence of the mirror symmetry in the $yz$-plane ($\lambda = 0$). 
(a) Two Weyl nodes on the $z$-axis for all $0 < \theta < \pi/2$. (b) Nodal line at $\theta = \pi/2$. This is a critical state, at which chiralities of the two Weyl nodes on the $z$-axis change signs. (c) Weyl nodes have exchanged 
chiralities for $\pi/2 < \theta < \pi$.}
\label{fig:1}
\end{figure}

To see what happens in detail, let us find how the spectrum of the Dirac Hamiltonian, perturbed by the applied magnetic (Zeeman) field
\beq
\label{eq:16}
H(\bk) = v_F (- \tau^z \sigma^y k_x + \tau^z \sigma^x k_y + \tau^y k_z) + b \cos \theta \sigma^z  + b \sin \theta \sigma^x, 
\eeq
where $b = g \mu_B B$, evolves as a function of the angle $\theta$. 
Diagonalizing Eq.~\eqref{eq:16}, one obtains the eigenstate energies
\beq
\label{eq:17}
\epsilon_{sr}(\bk) = s \sqrt{v_F^2 (k_x^2 + k_y^2 \cos^2 \theta) + m_r^2(\bk)},
\eeq
where $s, r = \pm$ and 
\beq
\label{eq:18}
m_r(\bk) = b + r v_F \sqrt{k_y^2 \sin^2 \theta + k_z^2}. 
\eeq
For any $\theta \neq \pi/2$, the two $r = -$ bands touch at two Weyl points, located on the z-axis at $k_z = \pm b/v_F$. 
The Fermi velocity, characterizing the dispersion away from the two points is, however, anisotropic:
\beq
\label{eq:19}
v_{Fx} = v_{Fz} = v_F, \,\, v_{Fy} = v_F |\cos \theta|.
\eeq 
When $\theta = \pi/2$, $v_{Fy}$ vanishes, and the $r = -$ bands touch along a nodal line in the $yz$-plane, given by the 
equation
\beq
\label{eq:20}
k_y^2 + k_z^2 = b^2/v_F^2. 
\eeq
The nodal line is protected by the emergent mirror symmetry with respect to reflections in the $yz$-plane and describes 
a critical state, at which the chiralities of the two Weyl points at $k_z = \pm b/v_F$ interchange as the magnetic field is rotated 
through the mirror-symmetric angle $\theta = \pi/2$, see Fig.~\ref{fig:1}. 

Now suppose the Fermi energy coincides with the Dirac point in the absence of the magnetic field, i.e. $\epsilon_F = 0$. 
The anomalous Hall conductivity as a function of the angle will then have the following form
\beq
\label{eq:21}
\sigma_{xy} (\theta) = \frac{e^2}{h} \frac{2 b /v_F}{2 \pi} \textrm{sign}(\cos \theta).
\eeq
This equation bears close resemblance to the equation for the Hall conductivity of a massive 2D
Dirac fermion of mass $m$ 
\beq
\label{eq:22}
\sigma_{xy} = \frac{e^2}{2 h} \textrm{sign}(m). 
\eeq
Eq.~\eqref{eq:22} expresses the parity anomaly of 2D Dirac fermions~\cite{Semenoff84,Haldane88}, namely the property 
that when $m \rightarrow 0$ from above or below, the Hall conductivity does not vanish, which apparently contradicts 
the time reversal and parity symmetry of the massless 2D Dirac Hamiltonian. 
In our case, the Hall conductivity does not vanish as the angle $\theta$ approaches the mirror-symmetric value $\theta = \pi/2$ from 
above or from below, although exactly at $\theta = \pi/2$ the Hall conductivity must vanish by symmetry. 

This resemblance to the parity anomaly is not accidental and may be understood as follows. 
As is well known, the Weyl nodes, which exist in this system for all $\theta \neq \pi/2$, 
are monopoles of the Berry curvature, at which the Chern number, characterizing 2D sections of the
BZ, perpendicular to the line, connecting a pair of Weyl nodes, i.e. the $z$-axis in our case, changes by the 
Weyl node charge upon crossing its location. 
In our case at $\theta = \pi/2$ the two Weyl nodes interchange their topological charge, which means that the 
Chern numbers of 2D BZ slices, perpendicular to the $z$-axis, change sign everywhere. 
Such a sign change requires two 2D Dirac fermions changing sign of their mass simultaneously at $\theta = \pi/2$ at 
every value of $k_z$ in between the Weyl node locations. 
These two massless 2D Dirac fermions are located at the intersections of the line node with fixed-$k_z$ sections of the BZ. 
We note that similar ideas have also been proposed in relation to Dirac nodal lines in PT-symmetric materials in Ref.~\cite{Schnyder17}. 

\begin{figure}[t]
\includegraphics[width=8cm]{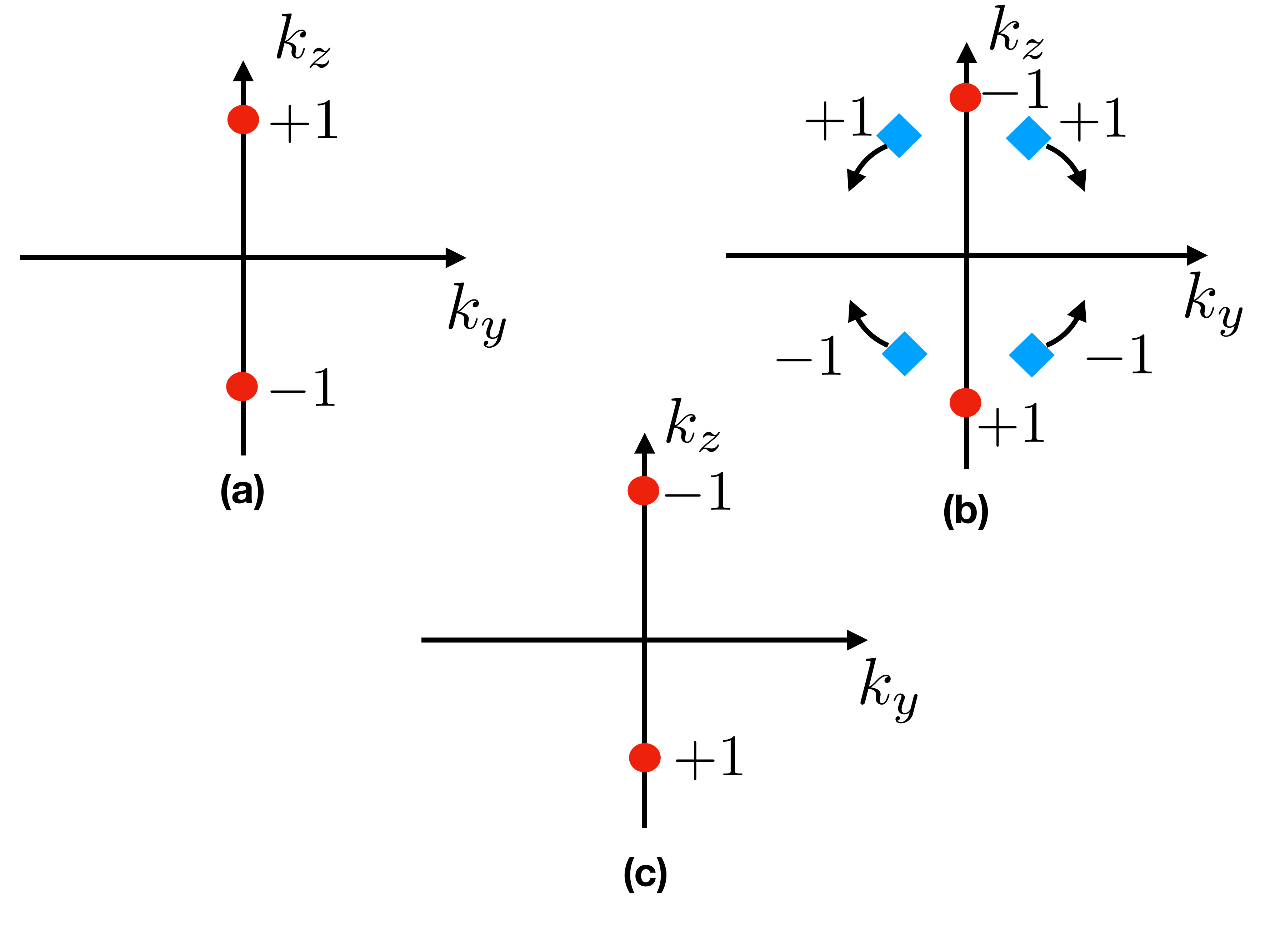}
\caption{(Color online) Evolution of the Weyl nodes without the mirror symmetry in the $yz$-plane ($\lambda > 0$). 
(a) Two Weyl nodes on the $z$-axis for $0 < \theta < \theta_c$. (b) Two additional pairs of Weyl nodes split off from the $z$-axis and move towards the $y$-axis for $\theta_c < \theta < \pi/2$. The two Weyl nodes on the $z$-axis change their chiralities 
at $\theta = \theta_c$. (c) The four additional Weyl nodes annihilate on the $y$-axis at $\theta = \pi/2$ and the two Weyl nodes with interchanged chiralities remain on the $z$-axis for $\pi/2 < \theta < \pi$.}
\label{fig:2}
\end{figure}
\begin{figure}[t]
\includegraphics[width=9cm]{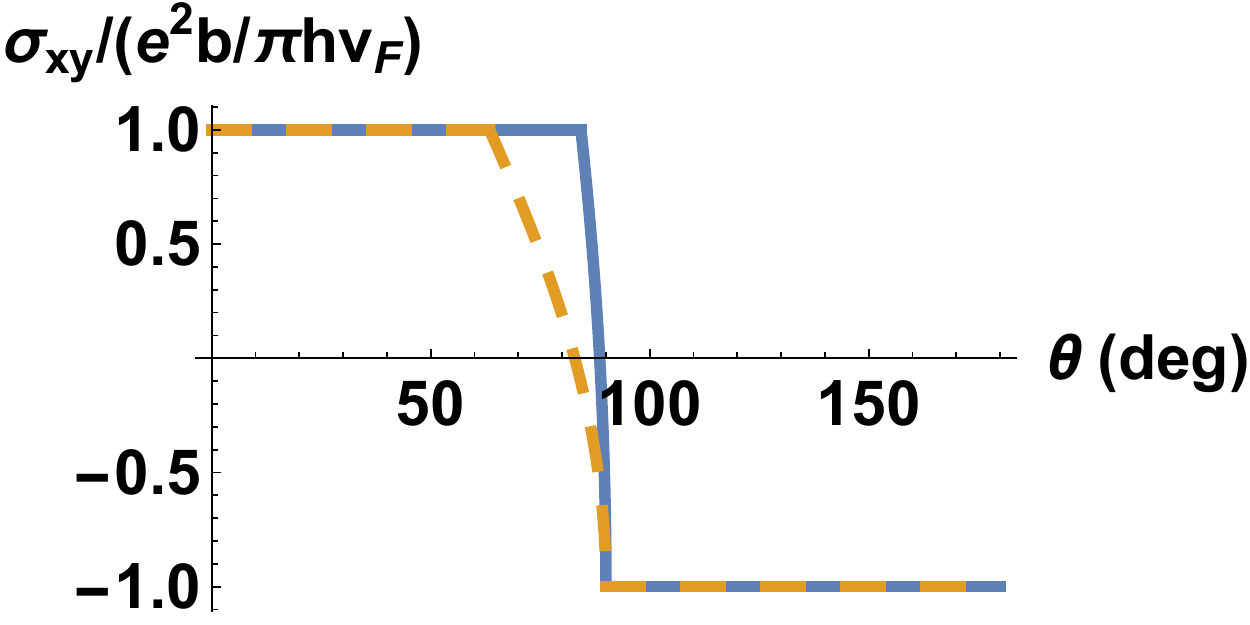}
\caption{(Color online) Anomalous Hall conductivity versus the angle $\theta$ between the magnetic field and the $z$-axis (the field is rotated in the $xz$-plane) for different values of the parameter $\delta = \lambda b^2/ 2 v_F^3$, which determines the degree of the 
mirror symmetry violation and broadening of the step-function singularity in $\sigma_{xy}$. The solid line corresponds to $\delta = 0.1$, while the dashed line to $\delta = 0.5$.}
\label{fig:3}
\end{figure}

So far we have analyzed the linearized Dirac Hamiltonian Eq.~\eqref{eq:16}. 
As explained above, such a linearized Dirac Hamiltonian possesses an emergent mirror symmetry in the $yz$-plane, 
which is what protects the nodal line at $\theta = \pi/2$. 
Let us now see what happens when we include the cubic term $d_5(\bk) \tau^z \sigma^z$, that violates this mirror 
symmetry (we will ignore the quadratic term $d_1(\bk) \tau^x$, since it does not violate mirror symmetry and thus does not lead to 
any qualitative changes). 
Eqs.~\eqref{eq:17} and \eqref{eq:18} are modified as 
\beqa
\label{eq:23}
&&\epsilon_{sr}(\bk) = \nonumber \\
&s&\sqrt{v_F^2 k_x^2 + [v_F \cos \theta - \lambda \sin \theta (k_z^2 - k_x^2)/2]^2 k_y^2 + m_r^2(\bk)}, \nonumber \\
\eeqa
and 
\beq
\label{eq:24}
m_r(\bk) = b + r \sqrt{v_F^2 k_z^2 + [v_F \sin \theta + \lambda \cos \theta (k_z^2 - k_x^2)/2]^2 k_y^2}. 
\eeq
The modified spectrum now does not have a nodal line for any values of the angle $\theta$. 
There is now always a pair of Weyl nodes on the z-axis at $k_z = \pm b/v_F$. 
In addition, there may exist four other Weyl nodes away from the $z$-axis in the $yz$-plane, whose location is 
given by
\beq
\label{eq:25}
k_z = \pm \frac{b}{v_F}\sqrt{\frac{\cot \theta}{\delta}},
\eeq
and 
\beq
\label{eq:26}
k_y = \pm \frac{b \sin \theta}{v_F}\sqrt{1 - \frac{\cot \theta}{\delta}},
\eeq
where $\delta = \lambda b^2/ 2 v_F^3$ is a parameter that defines the degree of the mirror symmetry violation (we assume 
$\lambda > 0$). 
The extra Weyl points appear on the $z$-axis at $k_z = \pm b/v_F$ (i.e. they split off 
the two original Weyl nodes that stay on the $z$-axis) at a critical angle 
\beq
\label{eq:27}
\theta_c = \textrm{arccot}(\delta). 
\eeq
At this point the topological charges of the two $z$-axis Weyl nodes change signs. 
The extra four nodes then move away from the $z$-axis as $\theta$ is increased and mutually annihilate on the $y$-axis at 
$k_y = \pm b/v_F$ when $\theta = \pi/2$, see Fig.~\ref{fig:2}. 
The step-function singularity in the anomalous Hall conductivity, Eq.~\eqref{eq:21}, is then broadened as
\beqa
\label{eq:28}
\sigma_{xy} = \frac{e^2 b}{\pi h v_F}\left\{
\begin{array}{c} 
1,\,\, 0 \leq \theta \leq \theta_c  \\ 
2 \sqrt{\frac{\cot \theta}{\delta}} - 1,\,\,\theta_c \leq \theta \leq \pi/2  
\nonumber \\
-1,\,\, \pi/2 \leq \theta \leq \pi  
\end{array}
\right. ,\\
\eeqa
which is illustrated in Fig.~\ref{fig:3}. 
The magnitude of the broadening is determined by the parameter $\delta$, which, in principle, may be made arbitrarily small by decreasing the magnitude of the applied field. 
Even when broadened by a finite $\delta$, the dependence of the anomalous Hall conductivity on the angle $\theta$ in 
Eq.~\eqref{eq:28} is highly unusual. 
Indeed, one would normally expect the magnitude of $\sigma_{xy}$ to be determined by the out-of-plane component of the 
magnetic field, i.e. $b \cos \theta$, and be simply proportional to it at low fields. 
Instead, at low fields, $\sigma_{xy}$ is proportional to the total magnitude of the field $b$, except in a narrow interval 
$\theta_c < \theta < \pi/2$, in which the dependence on $\theta$ is nonanalytic. 

In conclusion, we have demonstrated that Dirac semimetals possess the mirror anomaly, which is distinct from the 
chiral anomaly, and which manifests in a singular response of the Dirac semimetal to an applied magnetic field.
While we have considered only the simplest model of a Dirac semimetal, with a single Dirac point at TRIM, we do not expect the results to change qualitatively in the presence of several symmetry-related Dirac points.
The effect we have described has some potential for technological application: the extreme sensitivity of $\sigma_{xy}$ to the direction 
of the applied field near the mirror-invariant angle $\theta=\pi/2$ suggests transistor-like action. 
  
\begin{acknowledgments}
We acknowledge useful discussions with G. Bednik. Financial support was provided by NSERC of Canada. 
\end{acknowledgments}
\bibliography{references}

\end{document}